# Optimising Maintenance: What are the expectations for Cyber Physical Systems

Erkki Jantunen
Smart Industry and Energy Systems
VTT Technical Research Centre of Finland
Espoo, Finland
erkki.jantunen@vtt.fi

Urko Zurutuza
Electronics and Computing Department
Mondragon University
Mondragon, Spain
uzurutuza@mondragon.edu

Luis Lino Ferreira
CISTER/INESC-TEC, ISEP
Polytechnic Institute of Porto
Porto, Portugal
llf@isep.ipp.pt

Pal Varga
BME-TMIT
Budapest University of Technology and Economics
Budapest, Hungary
pvarga@tmit.bme.hu

*Abstract*—The need for maintenance is based on the wear of components of machinery. If this need can be defined reliably beforehand so that no unpredicted failures take place then the maintenance actions can be carried out economically with minimum disturbances to production. There are two basic challenges in solving the above. First understanding the development of wear and failures, and second managing the measurement and diagnosis of such parameters that can reveal the development of wear. In principle the development of wear and failures can be predicted through monitoring time, load or wear as such. Monitoring time is not very efficient, as there are only limited numbers of components that suffer from aging which as such is the result of chemical wear i.e. changes in the material. In most cases the loading of components influences their wear. In principle the loading can be stable or varying in nature. Of these two cases the varying load case is much more challenging than the stable one. The monitoring of wear can be done either directly e.g. optical methods or indirectly e.g. vibration. Monitoring actual wear is naturally the most reliable approach, but it often means that additional investments are needed. The paper discusses how the monitoring of wear and need for maintenance can be done based on the use of Cyber Physical Systems.

## I. Introduction

The need for maintenance is based on the wear of components of machinery. If this need can be defined reliably beforehand so that no unpredicted failures take place the maintenance actions can be carried out economically with minimum disturbances to production. In real life, further preventive actions such as cleaning, painting, and lubrication are needed in order to avoid the wear of components of machinery. Still, for the purpose of keeping this study compact these are not discussed in this context and the focus is only in wear. There are two basic challenges in solving the above. First understanding the development of wear and failures, and second managing the measurement and diagnosis of such parameters that can reveal the development of wear. In principle the development of wear and failures can be predicted through monitoring time, load or wear as such.

### A. Monitoring Time (Time Based Maintenance)

Monitoring time is not very efficient as there are only limited numbers of components that suffer from aging which as such is typically the result of chemical wear i.e. changes in the material. In most cases the loading of components influences their wear. It should be noted that if a component of a machine suffers from wear that is caused by time only i.e. no other factors influence the wear, then the monitoring of wear would be easy. One could position a similar component of the same age outside the machine in a place where it could be easily seen. This kind of monitoring is not used in great extent, but surprisingly at the same time quite a lot of maintenance is carried out based on calendar i.e. Time Based Maintenance.

### B. Monitoring Load (Scheduled Maintenance)

In principle the loading can be stable or varying in nature. Of these two cases the varying load case is much more challenging than the stable one. In case the loading of the machinery stays constant when the machinery is running, the cumulative loading can be recorded by monitoring the running hours of the machinery. This example goes back to monitoring time and the maintenance actions can be scheduled using the calendar i.e. Time Based Maintenance can be used. In the case of stable loading condition the statistical distributions of how frequently the parts of machinery fail due to wear can be collected and used to help the timing of maintenance actions. In case the machinery is used at different loading levels the situation becomes much more complex. The main problem is that wear can be a linear function of loading but in many cases it is not. This means that doubling the load does not necessarily cause wear at two times the original wear rate. Instead doubling the load might make the wear rate increase by an order of magnitude i.e. ten times higher. Consequently, the nature of the wearing process must be known i.e. what causes it, and how does the increasing of loading influence it. For the definition of cumulative loading and through that cumulative wear it is necessary to record the used time at different load levels, and then translate this to an estimation of cumulative wear. It should be noted that the loading can be caused by some force but also by chemicals etc. Furthermore it should be noted that the monitoring of the end product i.e. how much has been produced or transferred etc. can reveal the cumulative loading in a very simple and reliable way. In case of varying load the use of statistical failure data should be used with care. Note, that it could be totally misleading in case the wear of components as a function of load level is not taken into account. For example, if a component can last for one year or forty years based on the load level, the average life time value of twenty years is meaningless from the view point of timing the maintenance actions.

*C. Monitoring Wear (Condition Based Maintenance)*

For the definition of need for maintenance, the monitoring of actual wear is naturally the most reliable approach, but it often means that additional investments are needed. The monitoring of wear can be done either directly e.g. using optical methods or indirectly e.g. relying on vibration measurements. By far the most reliable method to define whether some of the components of machinery are worn in such a way that they need maintenance is the use of direct methods to define how worn the components are. The above statement is self-evident and sounds perhaps a little bit silly. The main problem is that usually the components of machinery are hidden so that the measurement of wear could be very expensive since for the definition of wear the machine would have to be stopped and partially disassembled. Naturally if the machine stops due to the failure of one of its components, this is a clear and reliable indication of extensive wear. Since the direct monitoring of wear often is difficult and expensive, a number of condition monitoring methods are used for the definition of wear of components. Examples of such methods are vibration, acoustic emission, and temperature measurements. Naturally, the indirect methods are not as accurate as the direct methods, but typically their use is much less expensive since the machines don't have to be disassembled. However, there are costs related to condition monitoring, and that has limited its use i.e. the savings in maintenance costs and the income due to better availability and quality of production have to be higher than the cost of condition monitoring. In this first chapter the motivation and principles of defining the wear of the components of machinery have been presented. Clearly in order to improve the availability of machinery there is a high need to carry out measurements in a competent way to support the definition of the need to carry out maintenance in proficient way. In chapter two the role of CPS in improving the efficiency of maintenance is discussed covering issues related to sensors, signal analysis, diagnosis of wear, and prognosis of wear development. In chapter three the IoT platforms to support the data processing are discussed covering data collection and data processing aspects. In chapter four the focus is in describing the industrial use cases where the new developed methods and technologies will tested and used.

## II. CYBER PHYSICAL SYSTEMS

As explained in the introduction, condition monitoring and following the CBM strategy can be challenging. In principle a certain sequence has to be followed. The ISO 13374 standard [1] has documented this sequence quite nicely. The various steps are discussed more in detail in the following paragraphs. In this paper CPS are understood to really cover this part i.e. they are systems that can measure and process information up to the level that makes it usable and understandable for the end user.

*A. Sensors*

Cheap, reliable, and technically good sensors are the key to the successful use of CBM. The recent development of sensor technology i.e. MEMS technology that has enabled the pre-processing of signals at an affordable price level has really made an enormous change. It is worth noticing that the use of wireless technology has increased remarkably, and that a lot has taken place in the development of sensors that have low energy consumption or can harvest the energy from the environment (light, vibration etc.). As such, condition monitoring is a very demanding area, which relies on a large set of sensor types, which can classified as: Native Equipment sensors, Maintenance Specific sensors and Maintenance Software sensors. Native Equipment sensors are required for the normal operation of specific equipment and used for control, monitoring, and security functionalities. But, in most situations these sensors can also be used for maintenance purposes, e.g. to detect the imbalance of power on an industrial machine. Maintenance Specific sensors are those sensors whose main purpose is to do condition monitoring of specific equipment parts, e.g. the wear of a machine part. It is also common to have Maintenance Software sensors, which combine and correlate the measurements of multiple sensors over time to indirectly determine or estimate a possible problem with the equipment. From multiple scenarios available in the literature [2] and from the pilots of transnational projects related with maintenance [3, 4] it is possible to identify the most used sensors. Temperature is probably the most used sensor type. Basically, such sensors are capable of measuring the temperature of fluids and gases (like water, oil, air or cooling liquids) and solids (like specific parts of a machine). Such measurements can be done directly (on the surface or in the liquid or gas) or indirectly (e.g. by measuring infrared emissions). Condition monitoring also relies on techniques to analyse the variations of temperature over time and infer any possible failure which causes an abnormal behaviour on the temperature. It is normal to have these sensors immediately available since they are cheap and can be easily be integrated into any installation. There are different types of pressure sensors with interest from condition monitoring point of view. It is possible to find these sensors being used on multiple industrial machines to determine the pressure of fluids, and to determine the load the machine is applying to the parts being manufactured. As an example, these sensors can be used to detect if there are leaks in hydraulic systems. Oil quality sensors are mostly specific to condition monitoring since they are capable of determining the degree of impurities in oil. Different types of sensors are available, alone or combined in the same device, capable of determining if the oil contains water, glycol, ferrous impurities, etc. Ferrous impurities allow the detection of possible wear on hydraulic cylinders, which might trigger maintenance operations in order to repair or replace them. The cost of these sensors is high, due to their complexity. Geometry-related sensors are sensors which are capable of measuring the deformation of parts of a machine or installation e.g. laser sensors, fibre optical, cameras, etc. They can immediately be available for high precision machines, which are able to compensate deformations on their structure or they can be deployed specifically for maintenance-related purposes. They are used to determine if parts of an installation or machine have deformations which might compromise safety or are symptoms of a problem. Vibration sensors are nowadays available at low cost due to recent advances on MEMS devices. Their basic operating principle is to measure acceleration over three axes. They can be used to e.g. monitor the movement of parts and determine if there are deviations from the normal behaviour. There are several different types of wear sensors, which have very different operating principles adapted to their specific use-case. For example machine tools use tools which are very sensitive to wear. Cameras can be used to determine the wear of tools, allowing the changing of these tools only when it is really necessary, thus reducing downtime. Soft sensors rely on the measurements performed by several physical sensors so that the signals of these are combined together and analysed.

*B. Signal Analysis*

Quite identically to the sensors i.e. that they are technically sufficient it is of high importance that the signal analysis of the condition monitoring signals is carried out in a proper way. The main motivation in signal analysis is to reveal features in the signal that indicate the existence of a fault i.e. the motivation is to separate the meaningful information from noise. Another import aspect is the reduction of information that needs to be sent further up in the process. In case quasi-static measurements i.e. if the measured parameters don't change their value dynamically the signal analysis methods are most often limited to the calculation of statistical parameters such as the average value, standard deviation, etc. In the case of more dynamic signals such as mechanical vibration, more sophisticated analysis techniques are used such as Fast Fourier Transform (FFT). Even though FFT has been used widely since the sixties it is surprising how often dramatic mistakes are made during the analysis. Probably one of the most common mistakes is that the measured signal is not low pass filtered prior to FFT which enables the so called aliasing phenomenon to take place. Another common mistake is that the frequency range that is used in the data acquisition phase does not support the analysis i.e. the phenomena the analyser is interested in are outside the measurement range. This is something that probably takes place in every second laboratory test with rolling element bearings because the natural frequency of these very small bearings is so high that it falls outside the measurement range. The standard equipment that are used for analysing vibration signals, which are the most common condition monitoring signals, are very conservative in the sense that they usually don't include any other analysis functions besides the FFT. Consequently it is rather typical to use AD cards with some mathematical libraries for the data collection and analysis. Numerous studies have been published about potential new signal analysis methods for condition monitoring. Unfortunately in many cases the researchers have not compared their new method with the commonly used techniques which means that even though the new methods seem to work in laboratory there is no proof that they would make an improvement to the commonly used techniques. Covering this subject there are also a number of good books available; one example is that of Randall [5].

*C. Diagnosis*

The development of sensors and signal analysis techniques described in the previous paragraphs together with the reduction in price has led to the situation where much more information is available than has been the case previously. This in turn has created great challenges in order to be able to automate the diagnosis based on this data. With diagnosis we mean here in this context the capability of turning the measured data into meaningful information regarding the condition of the machinery or devices that are being monitored. Automated diagnosis of condition monitoring signals has during the last years been an area of research where quite a massive number of papers have been published. There also exist a number of books that are focused on this subject. To name a few there is the book by Williams et al. Condition-Based Maintenance and Machine Diagnostics [6] which even though originally published in 1994 is still very relevant since it covers all the basic elements: maintenance information system, basic diagnostic techniques, most important condition monitoring methods, and interestingly also the system approach to condition monitoring. Another book worth mentioning here is that of Jardine & Tsang Maintenance, Replacement and Reliability, Theory and Applications [7]. It is interesting that the approach to help maintenance decision making in this book is so very different from that of the book of Williams et al. as here the focus is much more in statistical approaches. The differences in the approaches of the two books mentioned above is actually something that can be seen also today i.e. some of the approaches start from the idea that wear has to be understood in order to be able to diagnose it and other approaches start from the idea that if the mathematical/statistical approach is clever enough it can automatically diagnose a faulty situation. There exists the ISO 13379 Condition Monitoring and diagnostics of machines – General guidelines on data interpretation and diagnostic techniques standard [8]. The standard goes thoroughly through the basic diagnostic process and is as such of very high value. There is no point in repeating the contents of the standard but in order to understand the complexity i.e. the challenges it is worth mentioning some of the most important issues which are also related to what has been presented earlier in this paper. According to the standard the two main approaches that can be used for diagnosing a machine are: numerical methods and knowledge-based methods. There are also two possible ap-proaches to fault models: faults/symptoms approach and causal three approaches.

*D. Prognosis*

The key to excellence in CBM is the capability of making predictions i.e. to be able give prognosis on how long the component in questions will last in such a condition that doesn't stop the machine or cause any other harm. As such the task of making reliable prognosis is very demanding. According to the ISO 13381 [9] standard in the prognosis process one needs: to define the end point (usually the trip set point), to establish current severity, to determine/estimate the parameter behaviour and the expected rate of deterioration, and to determine the estimated time to failure (ETTF). The ISO 13381 [9] standard gives good advice of what needs to be considered when prognosis is carried out. The best known model how failures will take place is the so called bath-tube curve shown in Fig 1. In reality the bath-tube curve is one of the six basic types of wear curves that are commonly recognised. Of these six models three are such that it is realistic to assume that with efficient condition monitoring the development of wear could be noticed and consequently if the wear mechanism is understood some kind of prediction would be possible to make [10]. The other three wear models are such that it is not realistic to expect any kind of noticeable indication of the upcoming failure. The above indicated reference describes this issue more in detail and gives some very rough estimates how popular each of these types is. In rotating machinery the most typical component that needs to be maintained and monitored is the rolling element bearing. The rolling element bearing is a component that wears following the bath-tube curve and as such can be a subject to prognosis. However, even though rolling element bearings are very common, extremely vital for the machines to work, and a lot of research has been done in order to be able to diagnose the upcoming failures early and to be able to give prognosis of the expected lifetime of these components, there is still a lot to do in order to be able to reach really reliable prediction that can be done early [11, 12, 13, 14]. In Figure 2 an example is given of the development of vibration acceleration signal of a rolling element bearing in laboratory is given [15]. There exist physical explanations how come the measured curve looks the way it looks [16], however, it would take too much space to explain in this paper but the point is that it might not be easy to give prognosis based on such a condition monitoring signal.

### III. IoT Platforms to Support the Data Processing

Since embedded sensors and computing devices get deployed and start to interact with humans and with each other, they are nowadays referred to as CPS such as smart cities, buildings, homes, cars, factories; the list keeps growing. These smart systems (CPS) are often tied to one communication infrastructure. When this infrastructure provides opportunity for anything to connect and work on, heterogeneous system-of-systems can evolve serving various, dynamically changing needs. The basic idea behind the Internet of Things (IoT) initiatives is to make the push of the communication of the various systems including CPS towards a common way exactly as nodes are interconnected over the Internet. This means a lot more than "every system should have an IP address". This approach allows access to all sorts of methods and technologies already common in the world of Internet, including advanced security methods, resource sharing, etc. and not to forget cloud-based solutions, and big data analytics.

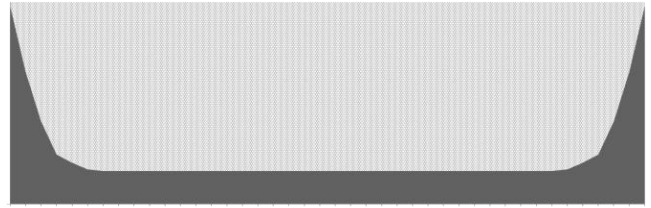

Fig. 1. Bath tube curve.

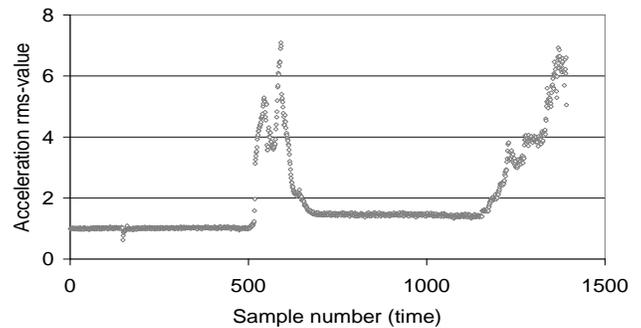

Fig. 2. Normalized vibration acceleration rms-value in laboratory tests with rolling bearing No. 3 [15].

*A. Collecting Sensor Data*

Without going into the details of communication infrastructures, we can assume that sensor data can be collected at a local level, at an aggregation point physically or in a network: logically close to those sensors. Depending on the level of maturity from the IoT point of view, even sensors can be equipped with the Internet Protocol (IP) stack. Although this simplifies interoperability and direct accessibility, it is not a critical requirement from CPS point of view. Sensor data can be propagated from the local aggregation point to a local proactive maintenance decision maker (especially in systems with real-time actuation requirements), or to a more sophisticated, high-level processing entity that has wider overview of the systems, their inter-relations, and surroundings.

*B. Data Pre-processing and Root Cause Analysis*

The data collected from the sensors should be correlated and if the patterns show unhealthy symptoms alarming actions should be acted upon. The generic data processing architecture is depicted in Figure 3. The relation of this architecture with the IoT view is the following. Deployed sensors or data sources can be either reported as standalone, remote entities, or through a local area network, e.g. part of a smart factory floor. This data and events can be correlated at a local aggregation point, based on locally available knowledge, such as sensor data, plant descriptions, or status in the workflow. On the other hand, this raw or pre-processed sensor data can be sent for further processing to humans (dashboard), state detection or health as-sessment processors, stream processors, or other, high-capability entities.

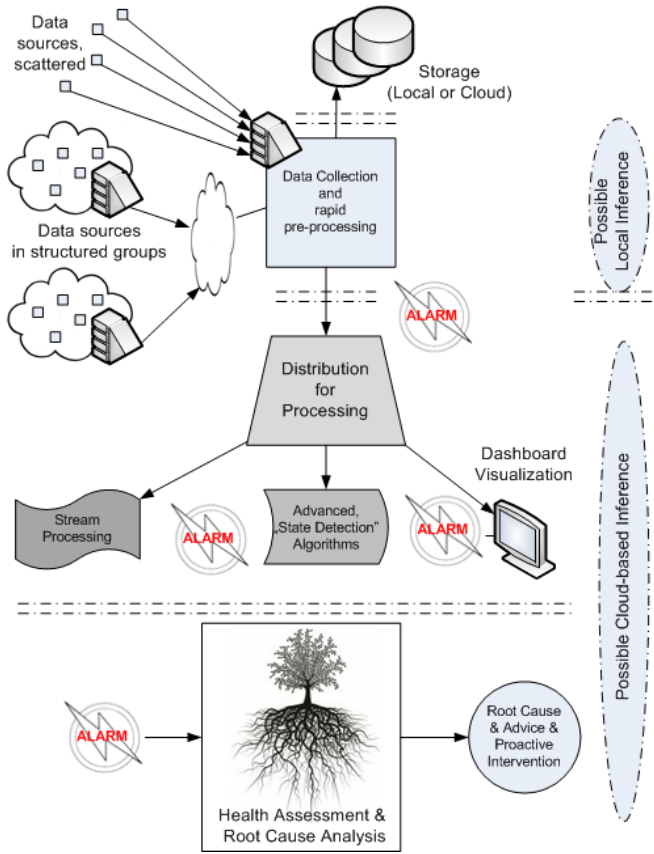

Fig. 3. The generic data processing architecture in Mantis [3] project.

On a low extreme, such a highly capable processor can reside in a central server whereas on a high extreme, it can be part of a cloud infrastructure, tuned for big data analytics. The alarms generated from the actual events or trending issues should be further analysed. When the root cause of the trend or the actual failure is found, the steps of intervention should be followed. Due to the wide range of popularity of the currently evolving big data analysis methods, central analysis of CPS sensor data is gaining momentum. The need of maintenance optimisation arises for very diverse sectors and scenarios (e.g., production asset maintenance, vehicle components, energy production systems, healthcare equipment maintenance). All of them have the common needs on using CPS to optimally monitor the wear of critical machine components, store, and aggregate maintenance, and production related information in order to discover and correlate it to aid decision-making mechanisms. In the Mantis [3] project the why (diagnosis), the when (prognosis), and the how, and by whom (optimal maintenance) are shared questions for the proposed use cases. Physical systems (e.g. industrial machines, vehicles, renewable energy assets) and the environment they operate in are monitored continuously. This is done by a broad and diverse range of intelligent sensors, resulting in massive amounts of data that characterise the usage history, operational condition, location, movement and other physical properties of those systems. These systems form part of a larger network of heterogeneous and collaborative systems (e.g. vehicle fleets or photovoltaic and windmill parks) connected via robust communication mechanisms able to operate in challenging environments.

IV. USE CASES

*A. Production Asset Maintenance Use Cases*

In a shaver production plant, Electro Chemical Machining (ECM) is used as a finishing technology for the mass production of shaver heads. It is a process that involves very detailed and precise tooling, which turns out to be delicate and expensive. This use case will evolve from a reactive maintenance strategy (maintenance is done after damage), to increase the predictability of the maintenance actions on the ECM processes, with particular focus on the wear of the tooling, hereby moving towards a data-driven proactive maintenance strategy. The process of pultrusion is a continuous, automatic and closed mould process. It is designed for high volume production at affordable costs. It consists of pulling through a high-temperature mould reinforcement impregnated with resin and a corresponding catalyst system. The current maintenance tasks within a pultrusion line are manual processes based on visual checks or timely programed tasks. In this use case, the need of developing new sensors is foreseen, while these will provide the basis for correlating the sensor data with environmental information and generate anomaly (wear) detection models. In a third use case, new sensors are being designed for understanding the behaviour of different components of a press machine. First, the early detection of fissures/cracks in form press machines' heads and caps; second, a thorough analysis and measurement of the force applications of the cranks, and third, monitoring the internal wear of a clutch-brake. This use case focuses on condition monitoring, whereas the optimization of the maintenance strategies might come after the manufacturers understand such behaviours. As another example, a full range of CPS as discussed in Section II is required by the owners of steel bending machine. Available measurements correlated with new sensor measurements, and data-driven fault diagnosis and prognosis are there to optimise current maintenance strategies. Finally, a leading compressor machine manufacturer will only need to exploit the data that they already monitor, and relate it with ERP, environmental data, or maintenance logs to determine the optimal maintenance interventions.

*B. Off-Road and Special Purpose Vehicles Use Cases*

Off-road and special purpose vehicles include trucks, buses, agricultural machinery, construction machinery and special vehicles (e.g. refuse collection vehicles, industrial trucks). These different vehicles have some common characteristics, which require the development of related technical solutions and technologies under technically similar challenges. Moreover, if detailed requirements to improve and optimize the maintenance of a high-performance materials handling equipment manufacturer are analysed, we would come up with an exact situation as in previous section; specific conditions of the vehicle components and liquids are to be monitored, for a later fault analysis and diagnosis, to finally produce decision-making mechanisms through a data-driven process. For this use case, hydraulic oil conditions, temperatures, load, pressures, and battery status will condition the wear analysis.

*C. Energy Production Asset Maintenance Use Cases*

Accelerometers and acoustic emissions are to be monitored in energy production assets for the proactive maintenance of wind turbines. The evolution of the acoustic signals over time, will inform about the current status of the wind turbine and its degradation process in relation to the initial and previous stage. Irradiance sensors are to be used to monitor and analyse the wear of photovoltaic plants components.

*D. Health Equipment Maintenance*

Healthcare Imaging Systems are essential for the diagnosis and treatment of patients in hospital and private clinics. Due to the large costs involved, it is not economically feasible to implement backup systems. Therefore, system uptime has to be maximized, planned downtime has to be minimized and unplanned shutdown has to be prevented. Every imaging system contains many sensors and generates large log files daily. Since these systems are heterogeneous by nature, the first challenge to address is to optimize logging such that data mining success (anamnesis). As a maintenance optimization strategy, rule-based proactive maintenance is foreseen. The rule engine shall input statistical parameters of the analytical model (e.g., precision, sensitivity, accuracy, MTTF/MTTB, remaining lifetime distributions etc.), life cycle and logistic costs and it should output decision rules about when and how to react proactively based on the outputs of the analytical models.

## V. CONCLUSION

In this paper the main principles of the monitoring of wear i.e. time-based, load-based, and wear-based have been presented. Following that the sequences of measurement, signal analysis, diagnosis, and prognosis have been described together with the challenges that are related to these phases. From these boundary conditions there follows a number requirements for the development of efficient CPS, which in turn could guarantee an optimal maintenance process. The presented debate is one stepping stone in the Mantis EU [3] project that has recently been started.

ACKNOWLEDGMENTS

The research has been conducted as a part of MANTIS Cyber Physical System based Proactive Collaborative Maintenance project. The project has received funding from the Electronic Component Systems for European Leadership Joint Undertaking under grant agreement No 662189. This Joint Undertaking receives support from the European Union's Horizon 2020 research and the national funding organisations Finnish Funding Agency for Innovation Tekes, Ministerio de Industria, Energía y Turismo (Spain) and by National Funds through FCT (Portuguese Foundation for Science and Technology) and by ERDF (European Regional Development Fund) through COMPETE (Operational Programme 'Thematic Factors of Competitiveness'), within project FCOMP-01-0124-FEDER-037281 (CISTER); by FCT and the ECSEL JU Grant nr. 662189 Call H2020-EE-2014-2015, and by the Hungarian National Research, Development and Innovation Office (NKFI), within project MANTIS, NEMZ_15-1-2016-0021.